\documentclass[twocolumn,english,pra]{revtex4}
\usepackage[T1]{fontenc}
\usepackage[latin9]{inputenc}
\usepackage{amsmath}
\usepackage{amssymb}
\usepackage{graphicx}
\usepackage{esint}

\makeatletter
\@ifundefined{textcolor}{}
{%
 \definecolor{BLACK}{gray}{0}
 \definecolor{WHITE}{gray}{1}
 \definecolor{RED}{rgb}{1,0,0}
 \definecolor{GREEN}{rgb}{0,1,0}
 \definecolor{BLUE}{rgb}{0,0,1}
 \definecolor{CYAN}{cmyk}{1,0,0,0}
 \definecolor{MAGENTA}{cmyk}{0,1,0,0}
 \definecolor{YELLOW}{cmyk}{0,0,1,0}
 }

\usepackage{babel}

\makeatother

\usepackage{babel}
\begin{document}

\title{Two-channel model description of confinement-induced Feshbach molecules}

\author{Shi-Guo Peng$^{1}$, Hui Hu$^{2}$, Xia-Ji Liu$^{2}$, and Kaijun
Jiang$^{1,3,}$}

\email{kjjiang@wipm.ac.cn}

\affiliation{$^{1}$State Key Laboratory of Magnetic Resonance and Atomic and
Molecular Physics, Wuhan Institute of Physics and Mathematics, Chinese
Academy of Sciences, Wuhan 430071, China }

\affiliation{$^{2}$ARC Centre of Excellence for Quantum-Atom Optics, Centre for
Atom Optics and Ultrafast Spectroscopy, Swinburne University of Technology,
Melbourne 3122, Australia}

\affiliation{$^{3}$Center for Cold Atom Physics, Chinese Academy of Sciences,
Wuhan 430071, China}

\date{\today}
\begin{abstract}
Using a two-channel model, we investigate theoretically the binding
energy of confinement-induced Feshbach molecules in two- and one-dimensional
ultracold atomic systems, near a Feshbach resonance. We show that
the two-channel prediction will evidently deviate from the simple
single-channel theory as the width of Feshbach resonances decreases.
For one-dimensional system, we perform a full two-channel calculation,
with the inclusion of bare interatomic interactions in the open channel.
Away from the resonance, we find a sizable correction to the binding
energy, if we neglect incorrectly the bare interatomic interactions
as in the previous work {[}Dickerscheid and Stoof, Phys. Rev. A \textbf{72},
053625 (2005){]}. We compare our theoretical results with existing
experimental data and present predictions for narrow Feshbach resonances
that could be tested in future experiments. 
\end{abstract}


\maketitle

\section{introduction}

The confinement-induced resonance (CIR) is one of the most intriguing
phenomena in low-dimensional ultracold atomic systems, and has attracted
a great deal of interest. It was first predicted by Olshanii using
a single-channel model in 1998 when considering two-body collisions
in one-dimensional (1D) harmonic waveguides \cite{Olshanii1998A}.
Later, this study was extended to two-dimensions by Petrov and co-workers
in 2000 \cite{Petrov2000B,Petrov2001I}. To date, there are a number
of experimental confirmations of the existence of CIR in both 1D and
2D setups.

To confirm experimentally CIR, it is convenient to use a Feshbach
resonance to tune the interatomic interactions \cite{Inouye1998O,Chin2010F}
and to measure the binding energy of the resulting confinement-induced
Feshbach molecules (CIFMs). The existence of CIFMs in a 1D Fermi gas
of $^{40}K$ atoms was first detected by Moritz \textit{et. al.} \cite{Moritz2005C}.
Using CIR in a 1D Bose gas of $^{133}Cs$ atoms \cite{Haller2009R},
Haller \textit{et. al.}\emph{ }realized the crossover from the Tonks-Girardeau
\cite{Girardeau1960R,Lieb1963E} to super-Tonks-Girardeau regime \cite{Astrakharchik2005B}.
Interestingly, in such 1D experiments \cite{Haller2010C}, an anomalous
splitting of the CIRs appeared when an anisotropy in the transverse
confinement was introduced, which could not be explained by generalizing
Olshanii's theory to anisotropic transverse confinement \cite{Peng2010C}.
But soon, it was realized that the splitting of the CIRs was resulted
from the coupling between the center-of-mass (COM) and relative motions
of two incoming atoms due to the anharmonicity of the trap \cite{Peng2011C,Sala2011I}.
Besides these 1D experiments, CIR was also recently observed in a
2D Fermi gas of $^{40}K$ atoms \cite{Frohlich2011R}. The interaction
between atoms was tuned by using a magnetic Feshbach resonance at
$B_{0}=224.2$G, and the binding energy of the CIFMs was measured.
Compared with the prediction of the single-channel theory \cite{Petrov2000B,Petrov2001I},
however, the observed binding energy is larger than the theoretical
prediction by 4 kHz in magnitude \cite{Frohlich2011R}. This discrepancy
may be understood from the picture of fermionic polaron, as suggested
by Schmidt \textit{et al.} \cite{Schmidt2012F}. Alternatively, the
discrepancy may also be solved by introducing an energy-dependent
interaction, as shown experimentally for $^{40}K$ atoms near another
Feshbach resonance $B_{0}=202.1$G \cite{Baur2012R}.

Here, we aim to study CIR and CIFMs by using a two-channel model,
with the inclusion of the effect of bound molecule states in the closed
channel. This is a physically more realistic description for Feshbach
resonances and therefore should provide a better description for CIFMs
which are measured experimentally. It is well-known that the two-channel
model yields the same theoretical predictions as the single-channel
model in the limit of broad Feshbach resonances \cite{Diener2005T,Liu2005S}.
The experimentally utilized resonances for $^{40}K$ and $^{133}Cs$
atoms are broad. As a result, our two-channel model description may
not give improved understanding of the existing CIR measurements.
However, our two-channel results should provide a useful guide for
future CIR experiments on relatively narrow Feshbach resonances.

We note that, a two-channel calculation was previously carried out
for 1D CIFMs by Dickerscheid and Stoof in 2005 \cite{Dickerscheid2005F}.
However, in their study the bare interatomic interactions were neglected.
It was shown then the two-channel model gives a better agreement with
the 1D experimental data than the single-channel model \cite{Dickerscheid2005F}.

In this work, we present a two-channel calculation for the binding
energy of CIFMs in 2D systems under an axially harmonic confinement.
Near the Feshbach resonance, the channel coupling will dominate the
contribution to the binding energy of the dressed molecules, which
allows us to neglect the bare interatomic interactions in the open
channel. At this resonance limit, we find that the two-channel result
coincides with the single-channel theory for a broad Feshbach resonance,
but will evidently deviates from the single-channel theory as the
resonance width decreases. We compare our theoretical calculation
with the recent 2D $^{40}K$ experiment \cite{Baur2012R}, in which
a broad Feshbach resonance at $B_{0}=202.1$G is used to control the
interatomic interactions. We find that the two-channel theory is in
agreement with the single-channel result, as expected, and both agree
well with the experimental data at the resonance. In order to demonstrate
the in-equivalence between the two-channel and single-channel theories,
we predict the binding energy of 2D CIFMs for two relatively narrow
resonances, $^{23}Na$ at $B_{0}=907$G and $^{87}Rb$ at $B_{0}=1007.4$G.

We also perform a \emph{full} two-channel calculation for 1D CIFMs.
We find that the bare interatomic interactions will give rise to a
sizable correction to the binding energy \emph{away} from Feshbach
resonances. This bare interatomic interaction was neglected in the
previous treatment \cite{Dickerscheid2005F}. By restoring the bare
interatomic interactions, the two-channel result coincides with the
single-channel prediction, as we anticipate for a broad resonance.
Thus, we realize that the previous better agreement claimed by Dickerscheid
and Stoof is not convincing.

The paper is arranged as follows. We first present the two-channel
Hamiltonian in Sec. \ref{sec:two-channel-hamiltonian}. Then the ansatz
of two-body wavefunctions for 2D and 1D systems is constructed in
Sec. \ref{sec:ansatz-of-the}. In Sec. \ref{sec:solution-of-two-body},
we solve the two-body problems and calculate the binding energy of
CIFMs. In Sec. \ref{sec:results-and-discussion}, we report the binding
energy as a function of the width of Feshbach resonances and show
how the two-channel result deviates from the single-channel prediction
as the resonance width decreases. We also compare our theoretical
results with the recent experiments and discuss in detail the correction
of the bare interatomic interactions to the binding energy of 1D CIFMs.
Finally, our main results are summarized in Sec. \ref{sec:conclusion}.

\section{two-channel hamiltonian\label{sec:two-channel-hamiltonian}}

For a two-component Fermi gas with atomic mass $m$, the two-channel
effective Hamiltonian that we consider includes the following single-particle
Hamiltonian and interaction Hamiltonian \cite{Drummond1998C,Kheruntsyan2000M},
\begin{multline}
\mathcal{H}_{0}=\sum_{\sigma}\int d^{3}\mathbf{r}\hat{\psi}_{\sigma}^{\dagger}\left(\mathbf{r}\right)\left[-\frac{\hbar^{2}}{2m}\nabla^{2}+V_{ext}\left(\mathbf{r}\right)\right]\hat{\psi}_{\sigma}\left(\mathbf{r}\right)\\
+\int d^{3}\mathbf{r}\hat{\Psi}^{\dagger}\left(\mathbf{r}\right)\left[-\frac{\hbar^{2}}{4m}\nabla^{2}+2V_{ext}\left(\mathbf{r}\right)+\Delta\left(B\right)\right]\hat{\Psi}\left(\mathbf{r}\right)\label{eq:T1}
\end{multline}
 and 
\begin{multline}
\mathcal{H}_{int}=U\int d^{3}\mathbf{r}\hat{\psi}_{\uparrow}^{\dagger}\left(\mathbf{r}\right)\hat{\psi}_{\downarrow}^{\dagger}\left(\mathbf{r}\right)\hat{\psi}_{\downarrow}\left(\mathbf{r}\right)\hat{\psi}_{\uparrow}\left(\mathbf{r}\right)\\
+g\int d^{3}\mathbf{r}\left\{ \hat{\Psi}^{\dagger}\left(\mathbf{r}\right)\hat{\psi}_{\downarrow}\left(\mathbf{r}\right)\hat{\psi}_{\uparrow}\left(\mathbf{r}\right)+h.c.\right\} .\label{eq:T2}
\end{multline}
 Here, $\hat{\psi}_{\sigma}\left(\mathbf{r}\right)\left(\sigma=\uparrow,\downarrow\right)$
and $\hat{\Psi}\left(\mathbf{r}\right)$ are the field operators of
fermionic atoms in the open (atomic) channel and of bosonic molecules
in the closed (molecular) channel, respectively. $V_{ext}\left(\mathbf{r}\right)$
is the external trapping potential, which could be described by a
harmonic trap as a good approximation. In order to realize a 2D system,
a tight axial confinement is applied experimentally, while the radial
(transverse) confinement is much weaker. Thus, the radial motion of
atoms is approximatively free. Theoretically, we can only consider
a tight axial confinement $V_{ext}\left(z\right)=m\omega_{\parallel}^{2}z^{2}/2$
along the $z$-direction. In a like manner, for a 1D system, the external
trapping potential can be chosen as $V_{ext}\left(\boldsymbol{\rho}\right)=m\omega_{\perp}^{2}\rho^{2}/2$,
which is in $x-y$ plane with $\rho^{2}=x^{2}+y^{2}$. $\Delta\left(B\right)$
is the detuning between atomic and molecular channels, which could
be tuned by a magnetic field near a Feshbach resonance. $U$ describes
the bare interaction between atoms with different spins in the atomic
channel, and $g$ denotes the coupling strength between the atomic
and molecular channels. We have used the constants $U$ and $g$ to
describe the coupling processes both for the bare interatomic interaction
and the formation of molecules, which is analogous to those used in
the pseudopotential theory. Consequently, a divergence in high energy
will appear. Thus, a regularization must be introduced as we shall
mention later.

\section{ansatz of the two-body wavefunction\label{sec:ansatz-of-the}}

In order to calculate the binding energy of CIFMs, let us solve a
two-body problem. In this paper, we are only interested in the tight-confinement
limit, $\hbar\omega_{\parallel,\perp}\gg k_{B}T$ , where $k_{B}$
is the Boltzmann constant and $T$ is the temperature. For an ultracold
Fermi gas with a pretty low temperature, almost all the atoms occupy
the lowest-energy (ground) state of the confined direction and can
not be excited to the higher energy levels during collisions. Due
to the harmonic confinement, the COM motion is completely decoupled
from the relative motion of two atoms. Therefore, the COM of the two
atoms will always stay in the ground state of the confined direction.
In addition, there is always a global translation invariance in the
untrapped directions, which means the COM momentum $\mathbf{K}$ in
these directions is a good quantum number. Thus, an ansatz of the
two-body wavefunction could take the following form, 
\begin{multline}
\left|\varphi_{2}\right\rangle =\left[\int d^{3}\mathbf{r}e^{i\mathbf{K}\cdot\boldsymbol{\rho}}\phi_{0}\left(z\right)\hat{\Psi}^{\dagger}\left(\mathbf{r}\right)\right.\\
+\int d^{3}\mathbf{r}_{1}d^{3}\mathbf{r}_{2}e^{i\mathbf{K}\cdot\left(\boldsymbol{\rho}_{1}+\boldsymbol{\rho}_{2}\right)/2}\phi_{0}\left(\frac{z_{1}+z_{2}}{2}\right)\\
\left.\times Q_{2}\left(\mathbf{r}_{1}-\mathbf{r}_{2}\right)\hat{\psi}_{\uparrow}^{\dagger}\left(\mathbf{r}_{1}\right)\hat{\psi}_{\downarrow}^{\dagger}\left(\mathbf{r}_{2}\right)\right]\left|0\right\rangle \label{eq:A1}
\end{multline}
 for a 2D system, and, 
\begin{multline}
\left|\varphi_{1}\right\rangle =\left[\int d^{3}\mathbf{r}e^{iKz}\phi_{00}\left(\boldsymbol{\rho}\right)\hat{\Psi}^{\dagger}\left(\mathbf{r}\right)+\right.\\
+\int d^{3}\mathbf{r}_{1}d^{3}\mathbf{r}_{2}e^{iK\left(z_{1}+z_{2}\right)/2}\phi_{00}\left(\frac{\boldsymbol{\rho}_{1}+\boldsymbol{\rho}_{2}}{2}\right)\\
\left.\times Q_{1}\left(\mathbf{r}_{1}-\mathbf{r}_{2}\right)\hat{\psi}_{\uparrow}^{\dagger}\left(\mathbf{r}_{1}\right)\hat{\psi}_{\downarrow}^{\dagger}\left(\mathbf{r}_{2}\right)\right]\left|0\right\rangle \label{eq:A2}
\end{multline}
 for a 1D system. Here $\left|0\right\rangle $ stands for the vacuum
state, and $Q_{D}\left(\mathbf{r}\right)$ is the relative wavefunction
of the two atoms in a {}``D''-dimensional system, which is spatially
antisymmetric, \emph{e.g.}, $Q_{2}\left(-\mathbf{r}\right)=-Q_{2}\left(\mathbf{r}\right)$.
$\phi_{00}\left(\boldsymbol{\rho}\right)$ and $\phi_{0}\left(z\right)$
are the ground states of 1D and 2D harmonic oscillators, respectively.

\section{solution of two-body problem\label{sec:solution-of-two-body}}

\subsection{In a 2D system}

Let us first consider an axially confined Fermi gas, where atoms can
move freely in the $x-y$ plane. By acting the two-channel Hamiltonian
(\ref{eq:T1}) and (\ref{eq:T2}) onto the two-body wavefunction ansatz
(\ref{eq:A1}), the Sch\"{o}rdinger equation $\mathcal{H}\left|\varphi_{2}\right\rangle =\varepsilon\left|\varphi_{2}\right\rangle $
is equivalent to the following set of equations (see the Appendix
for details), 
\begin{equation}
\left(-\frac{\hbar^{2}}{2\mu}\nabla^{2}+\frac{1}{2}\mu\omega_{\parallel}^{2}z^{2}\right)Q_{2}\left(\mathbf{r}\right)+\left[UQ_{2}\left(\mathbf{r}\right)+g\right]\delta\left(\mathbf{r}\right)=EQ_{2}\left(\mathbf{r}\right),\label{eq:S1}
\end{equation}
 
\begin{equation}
E=\Delta\left(B\right)+gQ_{2}\left(0\right),\label{eq:S2}
\end{equation}
 where the COM part has been separated out. $\mathbf{r}=\mathbf{r}_{1}-\mathbf{r}_{2}$
is the relative coordinate, $\mu=m/2$ is the reduced mass, and $E=\varepsilon-\hbar^{2}K^{2}/4m-\hbar\omega_{\parallel}/2$
is the relative energy. From Eq.(\ref{eq:S1}), the bound-state solution
can be written as, 
\begin{equation}
Q_{2}\left(\mathbf{r}\right)=-\left[UQ_{2}\left(0\right)+g\right]G_{E}^{(2)}\left(\mathbf{r},0\right),\label{eq:S3}
\end{equation}
 where $G_{E}^{(2)}\left(\mathbf{r},\mathbf{r}^{\prime}\right)$ is
the Green's function that satisfies, 
\begin{equation}
\left(-\frac{\hbar^{2}}{2\mu}\nabla^{2}+\frac{1}{2}\mu\omega_{\parallel}^{2}z^{2}-E\right)G_{E}^{(2)}\left(\mathbf{r},\mathbf{r}^{\prime}\right)=\delta\left(\mathbf{r}-\mathbf{r}^{\prime}\right).\label{eq:S4}
\end{equation}
 After some straightforward algebra \cite{Peng2011C}, the Green's
function $G_{E}^{(2)}\left(\mathbf{r},0\right)$ is easily obtained,
\begin{multline}
G_{E}^{(2)}\left(\mathbf{r},0\right)=\frac{e^{-z^{2}/2d_{\parallel}^{2}}}{2\pi^{3/2}d_{\parallel}^{3}\hbar\omega_{\parallel}}\\
\times\int_{0}^{\infty}dt\frac{\exp\left(\frac{\epsilon}{2}t-\frac{e^{-t}}{1-e^{-t}}\cdot\frac{z^{2}}{d_{\parallel}^{2}}-\frac{1}{t}\cdot\frac{\rho^{2}}{d_{\parallel}^{2}}\right)}{t\sqrt{1-e^{-t}}},\label{eq:S5}
\end{multline}
 where $d_{\parallel}=\sqrt{\hbar/\mu\omega_{\parallel}}$ is the
harmonic length, and $\epsilon=E/\hbar\omega_{\parallel}-1/2$ . This
integral representation of the Green's function (\ref{eq:S5}) is
valid only for $\epsilon<0$. Obviously, when $r\rightarrow0$, the
Green's function diverges as, 
\begin{eqnarray}
\lim_{r\rightarrow0}G_{E}^{(2)}\left(\mathbf{r},0\right) & \approx & \frac{1}{2\pi^{3/2}d_{\parallel}^{3}\hbar\omega_{\parallel}}\int_{0}^{\infty}dt\frac{\exp\left(-\frac{1}{t}\cdot\frac{r^{2}}{d_{\parallel}^{2}}\right)}{t^{3/2}},\nonumber \\
 & = & \frac{1}{2\pi d_{\parallel}^{2}\hbar\omega_{\parallel}}\cdot\frac{1}{r},\label{eq:S6}
\end{eqnarray}
 which results in an energy divergence. This is what we have anticipated
since we use a pseudopotential method to describe the coupling process
both for the interatomic interactions and for the formation of molecules.

In order to eliminate this singularity, the Green's function must
be regularized as follows, 
\begin{eqnarray}
\lim_{r\rightarrow0}\mathcal{G}_{\epsilon}^{(2)}\left(\mathbf{r},0\right) & = & \lim_{r\rightarrow0}\left[G_{E}^{(2)}\left(\mathbf{r},0\right)-\frac{1}{2\pi d_{\parallel}^{2}\hbar\omega_{\parallel}}\cdot\frac{1}{r}\right],\nonumber \\
 & = & \frac{1}{2\pi^{3/2}d_{\parallel}^{3}\hbar\omega_{\parallel}}\mathcal{F}_{2}\left(\epsilon\right),\label{eq:S7}
\end{eqnarray}
 where, 
\begin{equation}
\mathcal{F}_{2}\left(\epsilon\right)=\int_{0}^{\infty}dt\left[\frac{\exp\left(\epsilon t/2\right)}{t\sqrt{1-e^{-t}}}-\frac{1}{t^{3/2}}\right].\label{eq:S8}
\end{equation}
 Then, by substituting Eqs.(\ref{eq:S7}) and (\ref{eq:S8}) into
Eq.(\ref{eq:S3}), $Q_{2}\left(0\right)$ can be easily solved, 
\begin{equation}
Q_{2}\left(0\right)=-g\left[2\pi^{3/2}d_{\parallel}^{3}\hbar\omega_{\parallel}\mathcal{F}_{2}^{-1}\left(\epsilon\right)+U\right]^{-1}.\label{eq:S9}
\end{equation}
 Consequently, by combining Eqs.(\ref{eq:S2}) and (\ref{eq:S9}),
the binding energy $E_{B}$ of 2D CIFMs should satisfy the following
self-consistent equation, 
\begin{equation}
E_{B}=\Delta\left(B\right)-g^{2}\left[2\pi^{3/2}d_{\parallel}^{3}\hbar\omega_{\parallel}\mathcal{F}_{2}^{-1}\left(E_{B}/\hbar\omega_{\parallel}-1/2\right)+U\right]^{-1}.\label{eq:S10}
\end{equation}
 By solving this self-consistent equation, the binding energy can
be obtained.

\subsection{In a 1D system}

For a 1D system, atoms can only move freely along the axial direction,
while the radial motion is frozen. Operating on $\left|\varphi_{1}\right\rangle $
, Eq.(\ref{eq:A2}), with the Hamiltonian (\ref{eq:T1}) and (\ref{eq:T2}),
the Sch\"{o}rdinger equation is deduced to the following set of equations,
\begin{equation}
\left(-\frac{\hbar^{2}}{2\mu}\nabla^{2}+\frac{1}{2}\mu\omega_{\perp}^{2}\rho^{2}\right)Q_{1}\left(\mathbf{r}\right)+\left[UQ_{1}\left(\mathbf{r}\right)+g\right]\delta\left(\mathbf{r}\right)=EQ_{1}\left(\mathbf{r}\right),\label{eq:S11}
\end{equation}
 
\begin{equation}
E=\Delta\left(B\right)+gQ_{1}\left(0\right),\label{eq:S12}
\end{equation}
 where the relative energy is $E=\varepsilon-\hbar^{2}K^{2}/4m-\hbar\omega_{\perp}$.
After the similar derivation as that for the 2D case, a self-consistent
equation of the binding energy $E_{B}$ can easily obtained, 
\begin{equation}
E_{B}=\Delta\left(B\right)-g^{2}\left[2\pi^{3/2}d_{\perp}^{3}\hbar\omega_{\perp}\mathcal{F}_{1}^{-1}\left(E_{B}/\hbar\omega_{\perp}-1\right)+U\right]^{-1},\label{eq:S13}
\end{equation}
 where $d_{\perp}=\sqrt{\hbar/\mu\omega_{\perp}}$ and 
\begin{equation}
\mathcal{F}_{1}\left(\epsilon\right)=\int_{0}^{\infty}dt\left[\frac{\exp\left(\epsilon t/2\right)}{\sqrt{t}\left(1-e^{-t}\right)}-\frac{1}{t^{3/2}}\right].\label{eq:S14}
\end{equation}
 Then the binding energy of the 1D CIFMs can be solved using the self-consistent
equation (\ref{eq:S13}).

\section{results and discussion\label{sec:results-and-discussion}}

\subsection{At the resonance}

The single-channel theory is used to describe the universality for
a strongly-interacting Fermi gas near a broad Feshbach resonance \cite{Ho2004PRL,Hu2007NatPhys},
in which all other length scales become irrelevant except the average
interatomic distance. However, there is always an intrinsic length
$r_{*}$ for all Feshbach resonances, which is related to the resonance
width. It can be defined as \cite{Diener2005T}, 
\begin{equation}
r_{*}=\frac{\hbar^{2}}{2ma_{bg}\Delta\mu\Delta B},\label{eq:R1}
\end{equation}
 where $a_{bg}$ is the background scattering length in the open channel,
$\Delta B$ is the width of the resonance, and $\Delta\mu$ is the
magnetic moment difference of the molecular state with respect to
the threshold of two free atoms. For a broad resonance, this intrinsic
length is quite small due to the large resonance width, thus it is
reasonable that the single-channel theory gives a good description
of Feshbach resonances. However, when the resonance width becomes
narrow, the intrinsic length $r_{*}$ can not be ignored anymore,
and will obviously affect the properties of the resonance. Consequently,
the prediction of the two-channel theory, which is physically more
realistic, is expected to deviate from that of the single-channel
theory. In order to demonstrate this deviation, let us focus on the
situation right at the resonance, where the bare interatomic interaction
$U$ could be reasonably neglected, as we shall discuss in greater
detail later. The channel coupling strength $g$ is given by \cite{Dickerscheid2005F,Drummond2004C},
\begin{equation}
g=\hbar\sqrt{\frac{4\pi a_{bg}\Delta B\Delta\mu}{m}}=\frac{\hbar^{2}}{m}\sqrt{\frac{2\pi}{r_{*}}}.\label{eq:R2}
\end{equation}
 Combining with Eq.(\ref{eq:S10}), the binding energy of 2D CIFMs
predicted by the two-channel theory at the resonance can be written
as, 
\begin{equation}
\frac{\epsilon_{B}}{\hbar\omega_{\parallel}}=-\frac{d_{\parallel}}{4\sqrt{\pi}r_{*}}\mathcal{F}_{2}\left(\frac{\epsilon_{B}}{\hbar\omega_{\parallel}}\right)-\frac{1}{2},\label{eq:R3}
\end{equation}
 where $\epsilon_{B}=E_{B}-\hbar\omega_{\parallel}/2$ .

\begin{figure}
\includegraphics[width=1\columnwidth]{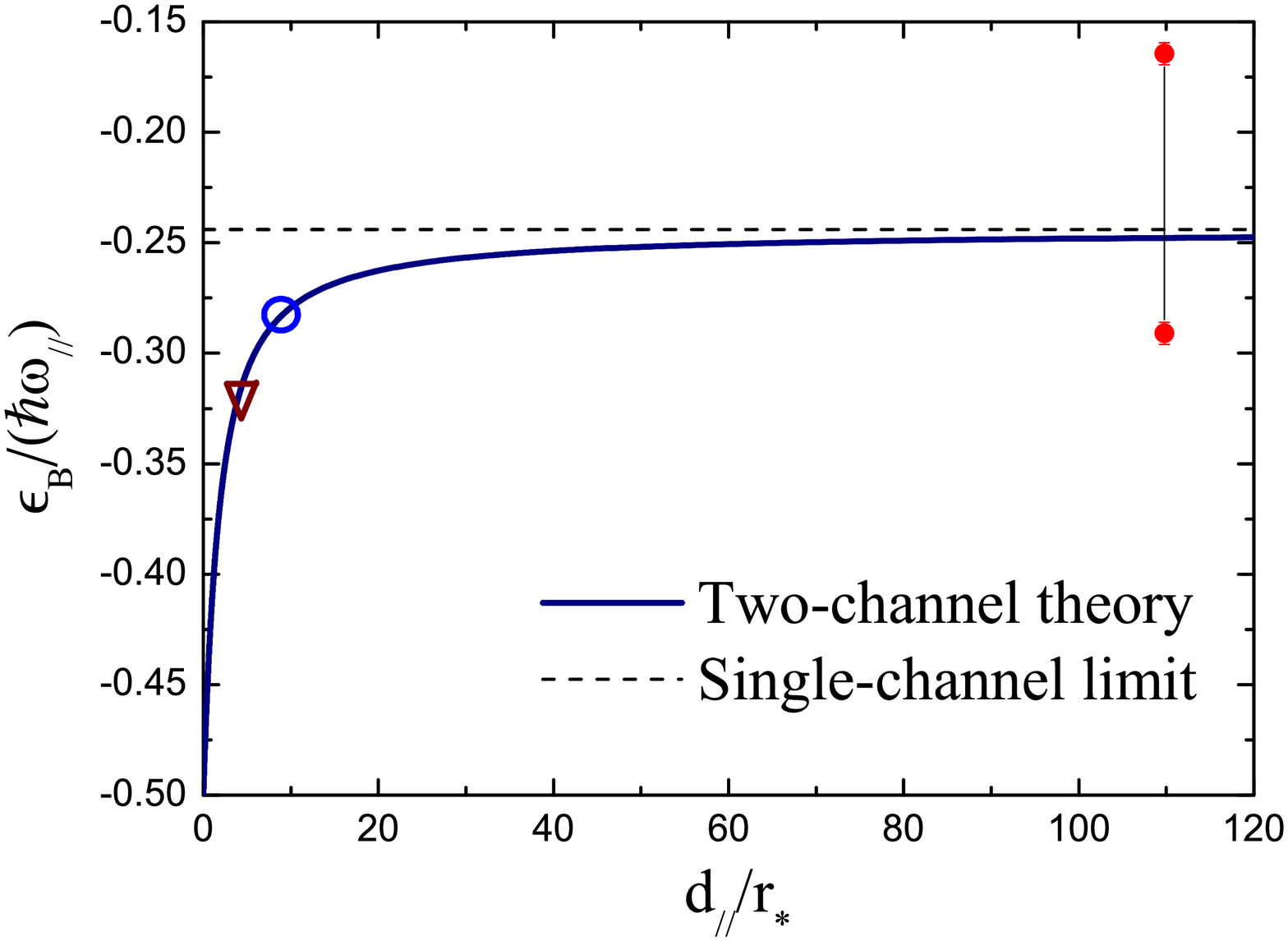}

\caption{(Color online) The binding energy of 2D CIFMs as a function of $d_{\parallel}/r_{*}$
right at the Feshbach resonance. The red solid circles with error
bar are taken from the 2D $^{40}K$ experiment \cite{Baur2012R} near
the Feshbach resonance at $B_{0}=202.1$G. The two-channel predictions
for the binding energy of two relatively narrow resonances for $^{23}Na$
at $B_{0}=907$G (blue circle) and $^{87}Rb$ at $B_{0}=1007.4$G
(wine triangle) are also presented. Here, the trap frequency $\omega_{\parallel}$
is chosen to be $2\pi\times75$ kHz as in Ref. \cite{Baur2012R}.}

\label{fig1} 
\end{figure}

The binding energy $\epsilon_{B}/\left(\hbar\omega_{\parallel}\right)$
of 2D CIFMs as a function of the intrinsic length $d_{\parallel}/r_{*}$
is shown in Fig.\ref{fig1}. At broad resonance limit, i.e., $d_{\parallel}/r_{*}\rightarrow\infty$
, the two-channel prediction approaches to the result of the single-channel
theory, which is predicted to be $-0.244\hbar\omega_{\parallel}$
\cite{Bloch2008M}. However, as the resonance becomes narrow, the
two-channel prediction will obviously deviate from the single-channel
result, and the binding energy of 2D CIFMs is strongly dependent on
the intrinsic length or the resonance width. In the recent 2D $^{40}K$
experiment \cite{Baur2012R}, the binding energy of 2D CIFMs is measured
across a broad Feshbach resonance at $B_{0}=202.1$G with $d_{\parallel}/r_{*}\sim110$.
The observed binding energy near the resonance is indicated in Fig.\ref{fig1}
by the red solid circles with error bar. Since there is no experimental
data exactly on the resonance, two points near the resonance are chosen,
and the binding energy right at the resonance should lie between these
two points.

We also predict the binding energy of 2D CIFMs for two relatively
narrow resonances of $^{23}Na$ at $B_{0}=907$G and $^{87}Rb$ at
$B_{0}=1007.4$G, which are presented by the blue circle and wine
triangle in Fig.\ref{fig1}, respectively.

\begin{figure}
\includegraphics[width=1\columnwidth]{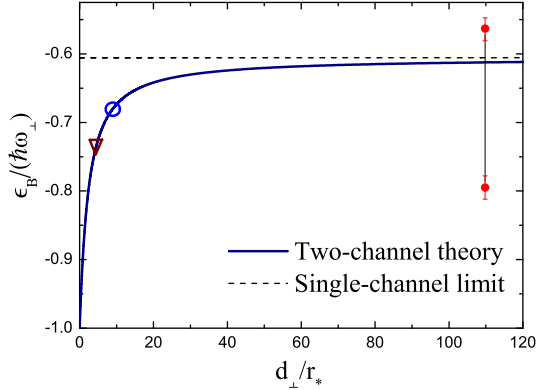}

\caption{(Color online) The same as Fig.\ref{fig1} but for 1D CIFMs. The red
solid circles with error bar are taken from the 1D $^{40}K$ experiment
\cite{Moritz2005C}, showing the two data points nearest to the Feshbach
resonance. Here, the trap frequency $\omega_{\perp}$ is chosen to
be $2\pi\times69$ kHz as in Ref. \cite{Moritz2005C}.}

\label{fig2} 
\end{figure}

In a like manner, the binding energy of 1D CIFMs at the resonance
is given by, 
\begin{equation}
\frac{\epsilon_{B}}{\hbar\omega_{\perp}}=-\frac{d_{\perp}}{4\sqrt{\pi}r_{*}}\mathcal{F}_{1}\left(\frac{\epsilon_{B}}{\hbar\omega_{\perp}}\right)-1
\end{equation}
where $\epsilon_{B}=E-\hbar\omega_{\perp}$ . The binding energy of
1D CIFMs as a function of the intrinsic length is plotted in Fig.\ref{fig2}.

\subsection{2D CIFMs across the Feshbach resonance}

In the following, let us investigate the binding energy of 2D CIFMs
across Feshbach resonances. By using Eqs. (\ref{eq:S8}) and (\ref{eq:S10}),
the binding energy predicted by the two-channel theory satisfies,
\begin{equation}
\epsilon_{B}+\frac{1}{2}\hbar\omega_{\parallel}=\Delta\left(B\right)-g^{2}\left[2\pi^{3/2}d_{\parallel}^{3}\hbar\omega_{\parallel}\mathcal{F}_{2}^{-1}\left(\frac{\epsilon_{B}}{\hbar\omega_{\parallel}}\right)+U\right]^{-1}.\label{eq:R4}
\end{equation}
 We recall that the binding energy given by the single-channel theory
is determined by the following equation \cite{Petrov2001I,Bloch2008M},
\begin{equation}
\frac{d}{a_{3D}}=-\frac{1}{\sqrt{\pi}}\mathcal{F}_{2}\left(\frac{\epsilon_{B}}{\hbar\omega_{\parallel}}\right),\label{eq:R5}
\end{equation}
 where $a_{3D}$ is the effective 3D s-wave scattering length.

In a magnetic Feshbach resonance, the detuning between the open and
closed channels, $\Delta\left(B\right)=\Delta\mu\left(B-B_{0}\right)$
, varies as a function of the magnetic field $B$ and depends on the
moment difference $\Delta\mu$ between the open and closed channels.
The bare interatomic interaction $U$ and the channel-coupling constant
$g$ are given by $U_{0}=4\pi\hbar^{2}a_{bg}/m$ and Eq.(\ref{eq:R2}),
respectively \cite{Dickerscheid2005F,Drummond2004C}. The effective
3D scattering length $a_{3D}$ is adjusted by the magnetic field $B$
according to, 
\begin{equation}
a_{3D}\left(B\right)=a_{bg}\left(1-\frac{\Delta B}{B-B_{0}}\right).\label{eq:R6}
\end{equation}

\begin{figure}
\includegraphics[width=1\columnwidth]{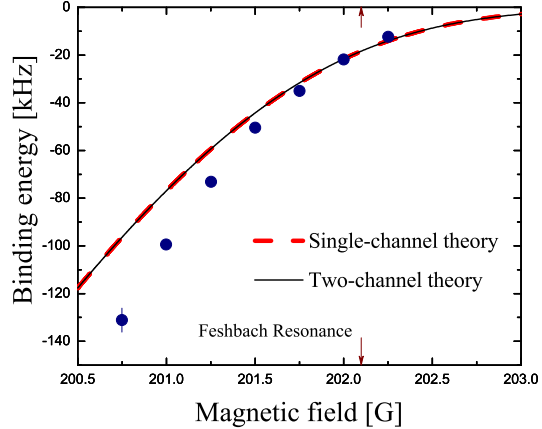}

\caption{(Color online) The binding energy of 2D CIFMs across the Feshbach
resonance at $B_{0}=202.1$G, comparing with the experimental data
(solid circles) \cite{Baur2012R}. The parameters for this resonance
are $\Delta B=7.8$G, $\Delta\mu=1.68\mu_{B}$, $a_{bg}=174a_{B}$,
where $\mu_{B}$ and $a_{B}$ are the Bohr magneton and radius, respectively.
The trap frequency in the strongly confined direction is $\omega_{\parallel}=2\pi\times75$
kHz.}

\label{fig3} 
\end{figure}

The two-channel prediction of the binding energy of 2D CIFMs across
a \emph{broad} Feshbach resonance as well as that of the single-channel
theory is illustrated in Fig.\ref{fig3}, comparing with the recent
2D $^{40}K$ experiment \cite{Baur2012R}. It can be seen clearly
that the two-channel theory agrees with the single-channel result
for such a broad resonance, and both agree well with the experiment
at the resonance. However, away from the resonance, the observed binding
energy deviates from both two-channel and single-channel predictions.

\subsection{1D CIFMs across the Feshbach resonance}

\begin{figure}
\includegraphics[width=1\columnwidth]{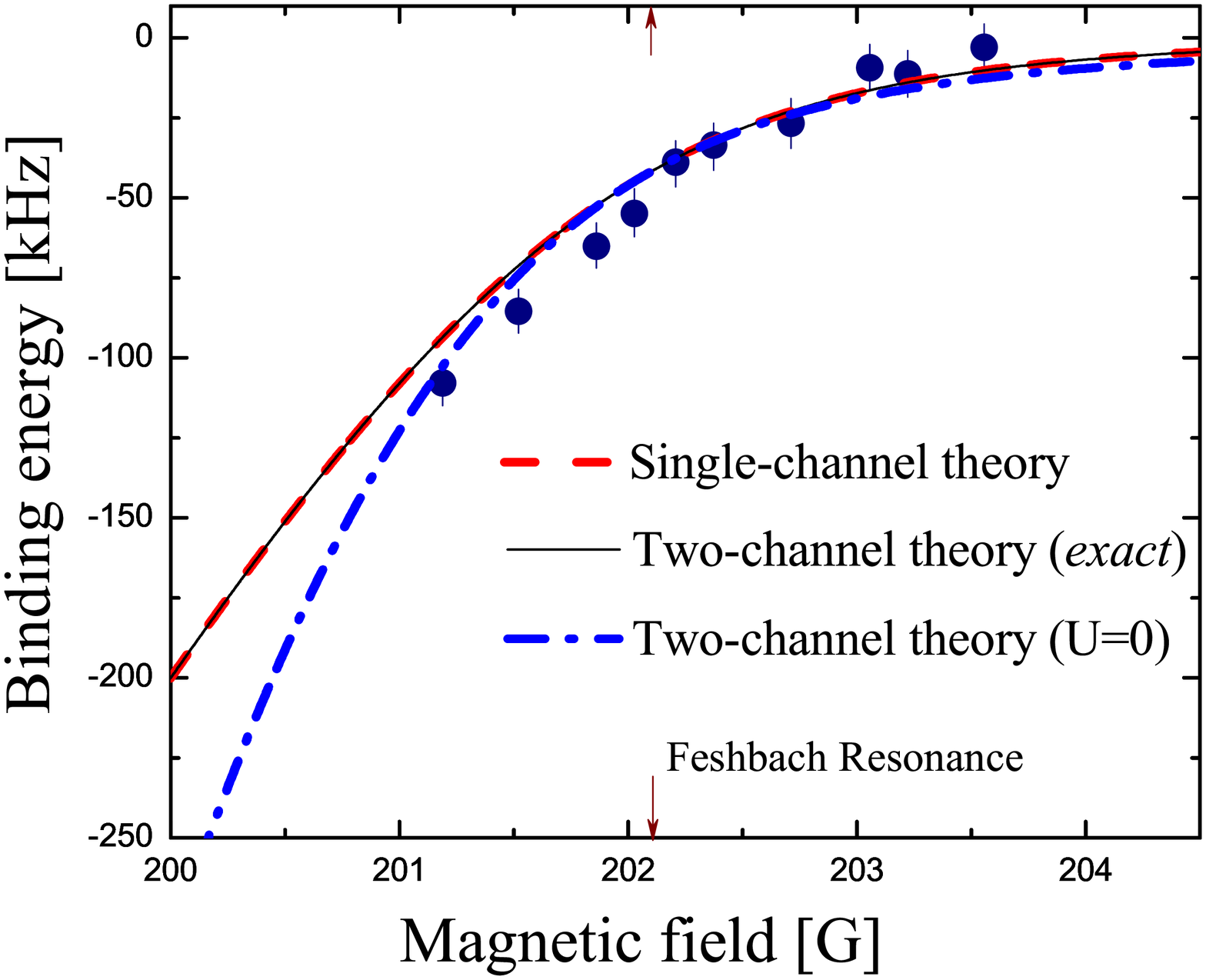}

\caption{(Color online) The binding energy of 1D CIFMs across the Feshbach
resonance at $B_{0}=202.1$G, comparing with the experiment data (solid
circles) \cite{Moritz2005C}. The blue dash-dotted curve is the result
of Ref. \cite{Dickerscheid2005F}, obtained by ignoring the bare interatomic
interaction in the open channel, i.e., $U=0$. Here, the trap frequency
in the strongly confined direction is $\omega_{\perp}=2\pi\times69$
kHz \cite{Moritz2005C}.}

\label{fig4} 
\end{figure}

We now turn to the binding energy of 1D CIFMs. According to Eqs.(\ref{eq:S13})
and (\ref{eq:S14}), the two-channel prediction for the binding energy
of 1D CIFMs is given by,
\begin{multline}
\epsilon_{B}+\hbar\omega_{\perp}=\Delta\left(B\right)-g^{2}\left[2\pi^{3/2}d_{\perp}^{3}\hbar\omega_{\perp}\times\right.\\
\left.\mathcal{F}_{1}^{-1}\left(\frac{\epsilon_{B}}{\hbar\omega_{\perp}}\right)+U\right]^{-1}.\label{eq:R6-7}
\end{multline}
 The two-channel result is plotted as well as the single-channel prediction
in Fig. \ref{fig4}. We find that the results of the two-channel and
single-channel theories are almost the same, and both agrees with
the experiment \cite{Moritz2005C} at the resonance. This agreement
is anticipated since the resonance used in \cite{Moritz2005C} is
also very broad, and the two-channel theory should approach the single-channel
result, as we already seen in Fig. \ref{fig2}.

In Ref. \cite{Dickerscheid2005F}, Dickerscheid and Stoof suggested
to ignore the bare interatomic interaction $U$, by assuming that
this background interaction is quite small comparing to the channel
coupling near the resonance. Under this approximation, Eq. (\ref{eq:R6-7})
yields, 
\begin{equation}
\epsilon_{B}+\hbar\omega_{\perp}\approx\Delta\left(B\right)-\frac{g^{2}}{2\pi^{3/2}d_{\perp}^{3}\hbar\omega_{\perp}}\mathcal{F}_{1}\left(\frac{\epsilon_{B}}{\hbar\omega_{\perp}}\right),\label{eq:R7}
\end{equation}
 which recovers the results in Ref. \cite{Dickerscheid2005F}. Here,
we define a different harmonic length $d_{\perp}=\sqrt{\hbar/\mu\omega_{\perp}}$
, which is $\sqrt{2}$ times larger than that in Ref. \cite{Dickerscheid2005F},
and $\mathcal{F}_{1}\left(\epsilon\right)=\sqrt{\pi}\zeta\left(1/2,-\epsilon\right)$
, where $\zeta\left(s,q\right)$ is the Hurwitz Zeta function. The
results of Eq. (\ref{eq:R7}) is plotted in Fig. \ref{fig4} by a
blue dash-dotted curve. It seems that the two-channel theory provides
an improved agreement with the experimental data, compared with the
single-channel theory. Here, however, we find that if the bare interatomic
interaction is correctly included, the result of the \emph{full} two-channel
theory returns back to that of the single-channel calculation, as
expected for a \emph{broad} Feshbach resonance.

Now, it is interesting to discuss the condition under which the bare
interatomic interaction $U$ could be neglected. From Eq. (\ref{eq:R6-7}),
we find whether the bare interatomic interaction could be neglected
lies on the competition between $2\pi^{3/2}d_{\perp}^{3}\hbar\omega_{\perp}\mathcal{F}_{1}^{-1}\left[\epsilon_{B}/\left(\hbar\omega_{\perp}\right)\right]$
and $U$, rather than the ratio of $g/U$. Thus, the condition can
be taken as, 
\begin{equation}
2\pi^{3/2}d_{\perp}^{3}\hbar\omega_{\perp}\mathcal{F}_{1}^{-1}\left(\frac{\epsilon_{B}}{\hbar\omega_{\perp}}\right)\gg U,\label{eq:R8}
\end{equation}
 or, 
\begin{equation}
\left|\mathcal{F}_{1}\left(\frac{\epsilon_{B}}{\hbar\omega_{\perp}}\right)\right|\ll\sqrt{\pi}\cdot\frac{d_{\perp}}{a_{bg}}.\label{eq:R9}
\end{equation}

\begin{figure}
\includegraphics[width=1\columnwidth]{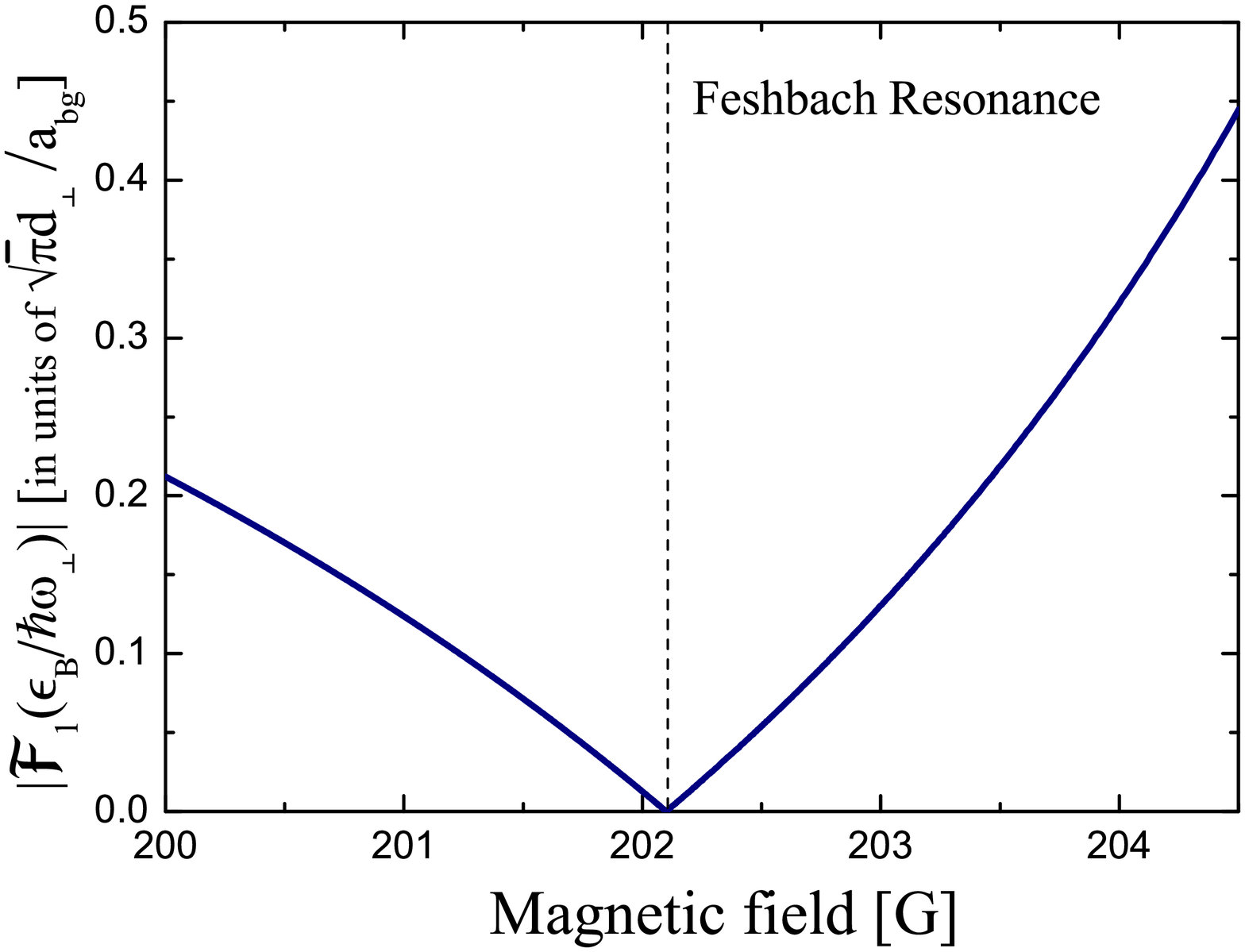}

\caption{(Color online) The function $\mathcal{F}_{1}\left[\epsilon_{B}/\left(\hbar\omega_{\perp}\right)\right]$
, in units of $\sqrt{\pi}d_{\perp}/a_{bg}$ , varies as a function
of the magnetic field for a 1D system across the resonance $B_{0}=202.1$G.
The dashed line indicates the location of the Feshbach resonance.}

\label{fig5} 
\end{figure}

The function $\mathcal{F}_{1}\left[\epsilon_{B}/\left(\hbar\omega_{\perp}\right)\right]$
as a function of the magnetic field is presented in Fig. \ref{fig5}.
Obviously, the condition (\ref{eq:R9}) can be fulfilled well very
close to the resonance, which means ignoring the bare interatomic
interaction is indeed a good approximation at the resonance limit.
However, away from the resonance, we see that the bare interatomic
interaction will bring a sizable correction to the binding energy
of the molecules, which makes the approximation results deviate from
those of the full two-channel calculations, as we have already seen
in Fig. \ref{fig4}.

\section{conclusions\label{sec:conclusion}}

In summary, we have presented a two-channel theory for calculating
the binding energy of confinement-induced Feshbach molecules in 2D
and 1D ultracold atomic systems. We have found that the two-channel
results will deviate from the single-channel predictions as the width
of Feshbach resonances decreases, although these two theories give
the same results in the limit of broad resonances. Compared with the
recent experiments, where a \emph{broad} magnetic Feshbach resonance
is used to control the interatomic interactions, both the two-channel
and single-channel theories agree well with the data near resonance.

For 1D confinement-induced Feshbach molecules, a full two-channel
calculation has been performed. Compared with the previous work \cite{Dickerscheid2005F},
in which the background interatomic interaction in the open channel
is neglected, we have found that the bare interatomic interaction
will yield a sizable correction to the binding energy away from the
resonance. Thus, in contrast to the difference shown in Ref. \cite{Dickerscheid2005F},
the two-channel result agrees very well with the single-channel prediction,
as we may anticipate for a \emph{broad} Feshbach resonance. 
\begin{acknowledgments}
Shi-Guo Peng and Kaijun Jiang are supported by China Postdoctoral
Science Foundation (Grant No. 2012M510187), the NSFC project (Grant
No. 11004224) and the NFRP-China (Grant No. 2011CB921601). Xia-Ji
Liu and Hui Hu are supported by the ARC Discovery Projects (Grant
No. DP0984637 and No. DP0984522) and the NFRP-China (Grant No. 2011CB921502). 
\end{acknowledgments}
\begin{widetext}

\appendix
%dummy comment inserted by tex2lyx to ensure that this paragraph is not empty

\section*{Appendix : The Sch\"{o}rdinger equation for a 2d system}

In this appendix, we present the detailed derivation of how the 2D
Sch\"{o}rdinger equation $\mathcal{H}\left|\varphi_{2}\right\rangle =\varepsilon\left|\varphi_{2}\right\rangle $
is reduced to the coupled equations (\ref{eq:S1}) and (\ref{eq:S2}).
By using the commutation and anti-commutation relations of the Bose
and Fermi field operators, 
\begin{equation}
\left[\hat{\Psi}\left(\mathbf{r}\right),\hat{\Psi}^{\dagger}\left(\mathbf{r}^{\prime}\right)\right]_{-}=\delta\left(\mathbf{r}-\mathbf{r}^{\prime}\right),\label{eq:APP1}
\end{equation}
 and 
\begin{equation}
\left[\hat{\psi}_{\sigma}\left(\mathbf{r}\right),\hat{\psi}_{\sigma^{\prime}}^{\dagger}\left(\mathbf{r}^{\prime}\right)\right]_{+}=\delta\left(\mathbf{r}-\mathbf{r}^{\prime}\right)\delta_{\sigma\sigma^{\prime}},\label{eq:APP2}
\end{equation}
 where $\mp$ represent the commutation and anti-commutation relations,
respectively, we obtain, 
\begin{eqnarray}
\mathcal{H}\left|\varphi_{2}\right\rangle  & = & \int d^{3}\mathbf{r}_{1}d^{3}\mathbf{r}_{2}\hat{\psi}_{\uparrow}^{\dagger}\left(\mathbf{r}_{1}\right)\hat{\psi}_{\downarrow}^{\dagger}\left(\mathbf{r}_{2}\right)\left\{ \left[-\frac{\hbar^{2}}{2m}\nabla_{\mathbf{r}_{1}}^{2}+V_{ext}\left(\mathbf{r}_{1}\right)-\frac{\hbar^{2}}{2m}\nabla_{\mathbf{r}_{2}}^{2}+V_{ext}\left(\mathbf{r}_{2}\right)+U_{1}\delta\left(\mathbf{r}_{1}-\mathbf{r}_{2}\right)\right]\times\right.\nonumber \\
 &  & \left.\times e^{i\mathbf{K}\left(\boldsymbol{\rho}_{1}+\boldsymbol{\rho}_{2}\right)/2}\phi_{0}\left(\frac{z_{1}+z_{2}}{2}\right)Q_{2}\left(\mathbf{r}_{1}-\mathbf{r}_{2}\right)+g\delta\left(\mathbf{r}_{1}-\mathbf{r}_{2}\right)e^{i\mathbf{K}\left(\boldsymbol{\rho}_{1}+\boldsymbol{\rho}_{2}\right)/2}\phi_{0}\left(\frac{z_{1}+z_{2}}{2}\right)\right\} \left|0\right\rangle +\nonumber \\
 &  & +\int d^{3}\mathbf{r}\hat{\Psi}^{\dagger}\left(\mathbf{r}\right)\left[\frac{\hbar^{2}K^{2}}{4m}+\frac{1}{2}\hbar\omega+\Delta\left(B\right)+gQ_{2}\left(0\right)\right]e^{i\mathbf{K}\cdot\boldsymbol{\rho}}\phi_{0}\left(z\right)\left|0\right\rangle .\label{eq:App3}
\end{eqnarray}
 Here, we have introduced pseudopotentials to describe the bare interatomic
interaction and the channel coupling. Then, by comparing the corresponding
terms in $\mathcal{H}\left|\varphi_{2}\right\rangle $ and $\varepsilon\left|\varphi_{2}\right\rangle $,
we arrive at, 
\begin{multline}
\left[-\frac{\hbar^{2}}{2m}\nabla_{\mathbf{r}_{1}}^{2}+V_{ext}\left(\mathbf{r}_{1}\right)-\frac{\hbar^{2}}{2m}\nabla_{\mathbf{r}_{2}}^{2}+V_{ext}\left(\mathbf{r}_{2}\right)+U\delta\left(\mathbf{r}_{1}-\mathbf{r}_{2}\right)\right]e^{i\mathbf{K}\left(\boldsymbol{\rho}_{1}+\boldsymbol{\rho}_{2}\right)/2}\phi_{0}\left(\frac{z_{1}+z_{2}}{2}\right)Q_{2}\left(\mathbf{r}_{1}-\mathbf{r}_{2}\right)\\
+g\delta\left(\mathbf{r}_{1}-\mathbf{r}_{2}\right)e^{i\mathbf{K}\left(\boldsymbol{\rho}_{1}+\boldsymbol{\rho}_{2}\right)/2}\phi_{0}\left(\frac{z_{1}+z_{2}}{2}\right)=\varepsilon e^{i\mathbf{K}\left(\boldsymbol{\rho}_{1}+\boldsymbol{\rho}_{2}\right)/2}\phi_{0}\left(\frac{z_{1}+z_{2}}{2}\right)Q_{2}\left(\mathbf{r}_{1}-\mathbf{r}_{2}\right),\label{eq:APP4}
\end{multline}
 and 
\begin{equation}
\left[\frac{\hbar^{2}K^{2}}{4m}+\frac{1}{2}\hbar\omega_{\parallel}+\Delta\left(B\right)+gQ_{2}\left(0\right)\right]e^{i\mathbf{K}\cdot\boldsymbol{\rho}}\phi_{0}\left(z\right)=\varepsilon e^{i\mathbf{K}\cdot\boldsymbol{\rho}}\phi_{0}\left(z\right).\label{eq:APP5}
\end{equation}
 If we seperate the COM motion from the relative motion, Eqs.(\ref{eq:APP4})
and (\ref{eq:APP5}) yield, 
\begin{equation}
\left(-\frac{\hbar^{2}}{2\mu}\nabla^{2}+\frac{1}{2}\mu\omega_{\parallel}^{2}z^{2}\right)Q_{2}\left(\mathbf{r}\right)+\left[UQ_{2}\left(\mathbf{r}\right)+g\right]\delta\left(\mathbf{r}\right)=EQ_{2}\left(\mathbf{r}\right),\label{eq:APP9}
\end{equation}
 and 
\begin{equation}
E=\Delta\left(B\right)+gQ_{2}\left(0\right),\label{eq:APP10}
\end{equation}
 respectively, where $\mathbf{r}=\mathbf{r}_{1}-\mathbf{r}_{2}$ ,
$\mu=m/2$ is the reduced mass, and $E=\varepsilon-\hbar^{2}K^{2}/4m-\hbar\omega_{\parallel}/2$
is the relative energy.

In a like manner, we can easily obtain Eqs.(\ref{eq:S11}) and (\ref{eq:S12})
for 1D systems.

\end{widetext}


\begin{thebibliography}{References}
\bibitem{Olshanii1998A}M. Olshanii, Phys. Rev. Lett. \textbf{81},
938 (1998).

\bibitem{Petrov2000B}D. S. Petrov, M. Holzmann, and G. V. Shlyapnikov,
Phys. Rev. Lett. \textbf{84}, 2551 (2000).

\bibitem{Petrov2001I}D. S. Petrov, and G. V. Shlyapnikov, Phys. Rev.
A \textbf{64}, 012706 (2001).

\bibitem{Chin2010F}C. Chin, R. Grimm, P. Julienne, and E. Tiesinga,
Rev. Mod. Phys. \textbf{82}, 1225 (2010).

\bibitem{Inouye1998O}S. Inouye, M. R. Andrews, J. Stenger, H.-J.
Miesner, D. M. Stamper-Kurn, and W. Ketterle, Nature (London) \textbf{392},
151 (1998).

\bibitem{Moritz2005C}H. Moritz, T. Stöferle, K. Günter, M. Köhl,
and T. Esslinger, Phys. Rev. Lett. \textbf{94}, 210401 (2005).

\bibitem{Haller2009R}E. Haller, M. Gustavsson, M. J. Mark, J. G.
Danzl, R. Hart, G. Pupillo, and H.-C. Nägerl, Science \textbf{325},
1224 (2009).

\bibitem{Girardeau1960R}M. Girardeau, J. Math. Phys. \textbf{1},
516 (1960).

\bibitem{Lieb1963E}E. H. Lieb and W. Liniger, Phys. Rev. \textbf{130},
1605 (1963).

\bibitem{Astrakharchik2005B}G. E. Astrakharchik, J. Boronat, J. Casulleras,
and S. Giorgini, Phys. Rev. Lett. \textbf{95}, 190407 (2005).

\bibitem{Haller2010C}E. Haller, M. J. Mark, R. Hart, J. G. Danzl,
L. Reichsöllner, V. Melezhik, P. Schmelcher, and H.-C. Nägerl, Phys.
Rev. Lett. \textbf{104}, 153203 (2010).

\bibitem{Peng2010C}S.-G. Peng, S. S. Bohloul, X.-J. Liu, H. Hu, and
P. D. Drummond, Phys. Rev. A \textbf{82}, 063633 (2010).

\bibitem{Peng2011C}S.-G. Peng, H. Hu, X.-J. Liu, and P. D. Drummond,
Phys. Rev. A \textbf{84}, 043619 (2011).

\bibitem{Sala2011I}S. Sala, P.-I. Schneider, and A. Saenz, e-print
arXiv:1104.1561.

\bibitem{Frohlich2011R}B. Fröhlich, M. Feld, E. Vogt, M. Koschorreck,
W. Zwerger, and M. Köhl, Phys. Rev. Lett. \textbf{106}, 105301 (2011).

\bibitem{Schmidt2012F}R. Schmidt, T. Enss, V. Pietilä, and E. Demler,
Phys. Rev. A \textbf{85}, 021602(R) (2012).

\bibitem{Baur2012R}S. K. Baur, B. Fröhlich, M. Feld, E. Vogt, D.
Pertot, M. Koschorreck, and M. Köhl, Phys. Rev. A \textbf{85}, 061604(R)
(2012).

\bibitem{Dickerscheid2005F}D. B. M. Dickerscheid and H. T. C. Stoof,
Phys. Rev. A \textbf{72}, 053625 (2005).

\bibitem{Diener2005T}R. B. Diener and T.-L. Ho, e-print arXiv:cond-mat/0405174.

\bibitem{Liu2005S}X.-J. Liu and H. Hu, Phys. Rev. A \textbf{72},
063613 (2005).

\bibitem{Drummond1998C}P. D. Drummond, K. V. Kheruntsyan, and H.
He, Phys. Rev. Lett. \textbf{81}, 3055 (1998).

\bibitem{Kheruntsyan2000M}K. V. Kheruntsyan and P. D. Drummond, Phys.
Rev. A \textbf{61}, 063816 (2000).

\bibitem{Ho2004PRL}T.-L. Ho, Phys. Rev. Lett. \textbf{92}, 090402
(2004).

\bibitem{Hu2007NatPhys}H. Hu, P. D. Drummond, and X.-J. Liu, Nature
Phys. \textbf{3}, 469 (2007).

\bibitem{Drummond2004C}P. D. Drummond and K. V. Kheruntsyan, Phys.
Rev. A \textbf{70}, 033609 (2004).

\bibitem{Bloch2008M}I. Bloch, J. Dalibard, and W. Zwerger, Rev. Mod.
Phys. \textbf{80}, 885 (2008).\end{thebibliography}
\end{document}